\documentstyle[emulateapj,10pt,apjfonts,psfig]{article}

\newcommand\psr{J2124--3358}
\newcommand\HI{H\,{\sc i}}

\newcommand\Ha{H$\alpha$}
\newcommand\kms{km~s$^{-1}$}
\newcommand\etal{{\rm et~al.\ }}

\newcommand\photunits{photons~s$^{-1}$~cm$^{-2}$}
\newcommand\lineSB{erg~s$^{-1}$~cm$^{-2}$~arcsec$^{-2}$}

\renewcommand\farcm{\hbox{$.\!\!^{\prime}$}}

\lefthead{GAENSLER, JONES \& STAPPERS}
\righthead{\uppercase{BOW SHOCK AROUND PSR~\psr}}

\begin{document}
\title{An Optical Bow Shock Around the Nearby Millisecond Pulsar \psr}
\submitted{Accepted to ApJ Letters 2002 Oct 24}
\author{B. M. Gaensler\altaffilmark{1}, D. H. Jones\altaffilmark{2, 3, 4},
and B. W. Stappers\altaffilmark{5, 6}}
\altaffiltext{1}{Harvard-Smithsonian
Center for Astrophysics, 60 Garden Street MS-6, Cambridge, MA 02138;
bgaensler@cfa.harvard.edu}
\altaffiltext{2}{European Southern Observatory, Casilla 19001, Santiago 19,
Chile}
\altaffiltext{3}{Observatorio Cerro Cal\'an, Departamento de
Astronom\'{\i}a, Universidad de Chile, Casilla 36-D, Santiago, Chile}
\altaffiltext{4}{Current address: Research School of Astronomy and
Astrophysics, Mount Stromlo Observatory, Cotter Road, Weston, ACT 2611,
Australia; heath@mso.anu.edu.au}
\altaffiltext{5}{Stichting ASTRON, 7990 Dwingeloo, The Netherlands}
\altaffiltext{6}{Sterrenkunding Institut ``Anton Pannekoek'', Kruislaan 403,
1098 SJ Amsterdam, The Netherlands; bws@science.uva.nl}
\hspace{-5cm}

\begin{abstract}

We report the discovery of an \Ha-emitting  bow-shock nebula powered by
the nearby millisecond pulsar \psr.  The bow shock is very broad, and
is highly asymmetric about the pulsar's velocity vector. This shape is
not consistent with that expected for the case of an isotropic wind
interacting with a homogeneous ambient medium.  Models which invoke an
anisotropy in the pulsar wind, a bulk flow of the surrounding gas, or a
density gradient in the ambient medium either perpendicular or parallel
to the pulsar's direction of motion also fail to reproduce the observed
morphology.  However, we find an ensemble of good fits to the nebular
morphology when we consider a combination of these effects. In all such
cases, we find that the pulsar is propagating through an ambient medium
of mean density $0.8-1.3$~cm$^{-3}$ and bulk flow velocity $\sim15-25$~\kms,
and that the star has recently encountered an increase in density by
$1-10$~cm$^{-3}$ over a scale $\la0.02$~pc. The wide variety of models
which fit the data demonstrate that in general there is no unique set
of parameters which can be inferred from the morphology of a bow-shock
nebula.

\end{abstract}


\section{Introduction}
\label{sec_intro}

Many pulsars experience significant ram pressure as they move through
the interstellar medium (ISM).  This pressure can confine the pulsar's
relativistic wind and generate a bow-shock pulsar wind nebula (PWN).
The \Ha\ emission generated by such sources provides an important probe
of the pulsar's interaction with its environment (see \cite{cc02} for a
recent review).  However, until recently only three \Ha\ PWNe were
known, and we consequently have lacked a detailed understanding of this
phenomenon.

We are carrying out a survey for new pulsar bow shocks in the southern
sky. We have previously reported the discovery of a faint \Ha\ PWN
surrounding PSR~B0740--28 (\cite{jsg02}). We here present another new
\Ha\ pulsar bow shock, associated with the nearby millisecond pulsar
\psr\ (\cite{bjb+97}).  A recent timing ephemeris (M.\ Bailes, private
communication) yields a spin period of 4.93~ms and a position (at MJD
50288.0) of RA (J2000) $21^{\rm h}24^{\rm m}43\fs86196(6)$,
Dec.\ (J2000) $-33^\circ58'44\farcs257(1)$, where the numbers in
parentheses indicate the uncertainty in the last digit.  The pulsar's
intrinsic spin-down luminosity is $\dot{E} =
4.3\times10^{33}$~erg~s$^{-1}$ (\cite{tsb+99}), while its dispersion
measure of 4.61~pc~cm$^{-3}$ implies a distance of $270\pm20$~pc
(\cite{cl02}). In future discussion, we adopt a distance $D =270d_0$~pc
to the source.

The pulsar timing ephemeris yields a proper motion relative to the
solar system barycenter of $52.6(3)$~milliarcsec~yr$^{-1}$ at position
angle $195\fdg6(3)$ (North through East).  Because of the pulsar's
proximity and low space velocity, the peculiar motion of the sun with
respect to the Local Standard of Rest needs to be
accounted for.  In the case of PSR~\psr, the sun's motion is directed
almost perpendicular to the line-of-sight, and so
contributes a significant component of the observed proper motion.
After correction for the solar peculiar velocity (and for the small
effects of differential Galactic rotation),
we find that the motion of
the pulsar relative to its local standard of rest is
47(3)~milliarcsec~yr$^{-1}$ at position angle $209(3)^\circ$,
corresponding to a projected space velocity of $(61\pm4)d_0$~\kms.  The
uncertainties in the direction and magnitude of the solar motion
dominate those in the pulsar's distance and measured proper
motion.

\section{Observations and Results}

The field around PSR~\psr\ was observed for 4800~sec on 2001 July 25
(MJD 52115) with a narrowband \Ha\ filter on the 3.5-m New Technology
Telescope (NTT) at the European Southern Observatory, La Silla, Chile.
The mean seeing was $\sim1\farcs2$; all other observational parameters
were as described by Jones \etal\ (2002\nocite{jsg02}).  Standard
techniques for bias and flat-field correction were used to remove
instrumental signatures from the data. Individual frames were dithered
to permit removal of CCD defects and cosmic rays using median filtering
on the frames when combined.  
An astrometric solution was determined using 10 field stars with
positions in the USNO Catalog.  The {\sc iraf} task {\sl ccmap} was
used to fit the positions with a third order polynomial with cross
terms. The RMS scatter was $(\Delta \alpha, \Delta \delta)$ = ($\pm
0\farcs15$,$\pm 0\farcs07$) and the residuals uniformly scattered. A
transformation of the derived USNO star coordinates back onto the image
did not reveal any systematic offsets between the observed
and derived positions.

\begin{figure}
\centerline{\psfig{file=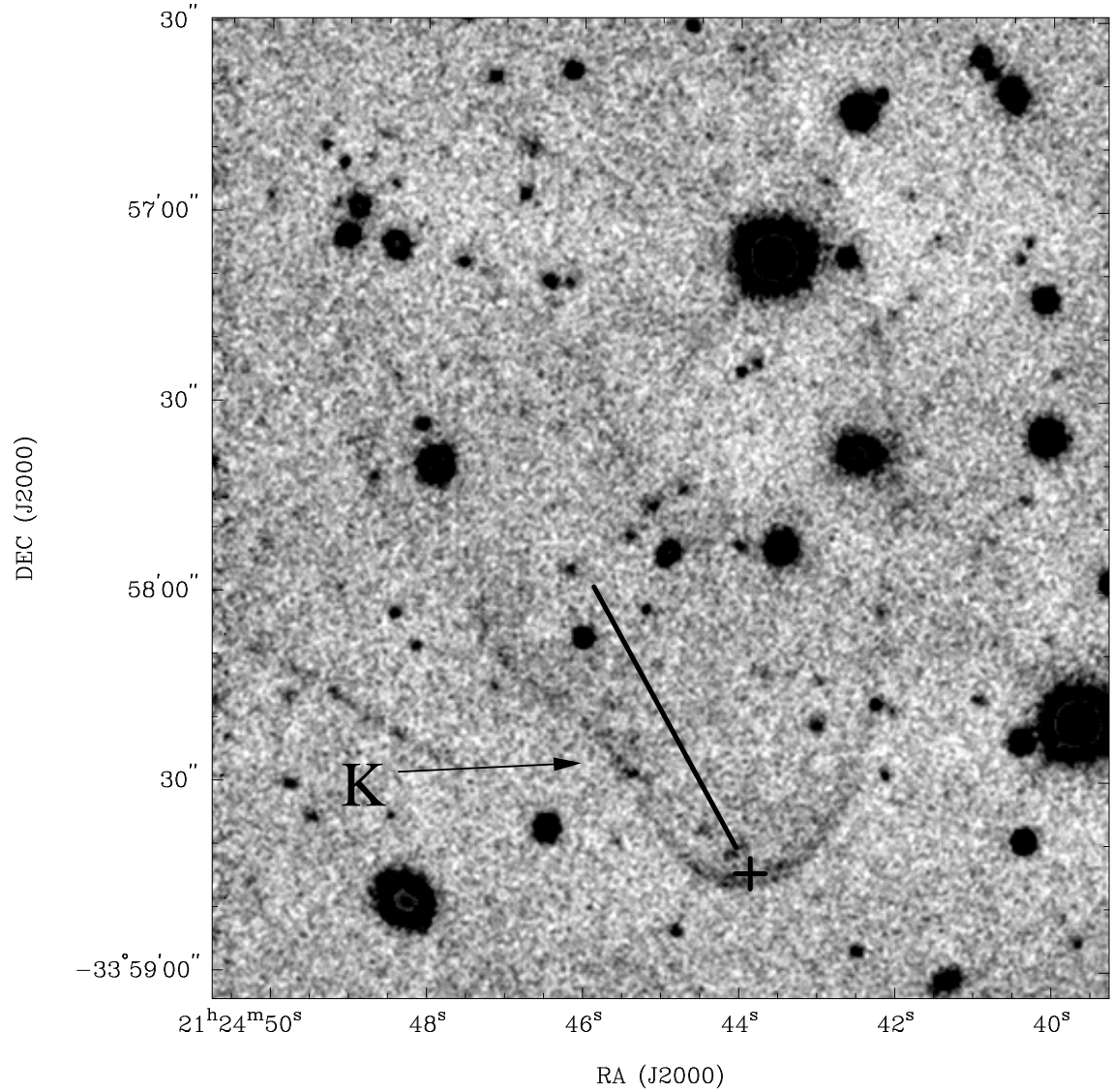,height=15cm}}
\caption{NTT \Ha\ image of PSR~\psr.  Counts within 40~ADU of the
nebular mean have been scaled by a factor of 10 and then box-car smoothed using a $5\times5$-pixel kernel.  The cross marks the position of the pulsar at
epoch MJD 52115, RA (J2000) $21^{\rm h}24^{\rm m}43\fs8563(1)$,
Dec.\ (J2000) $-33^\circ58'44\farcs511(3)$, while the straight line
indicates the pulsar's direction of motion in its local
standard of rest.
The length of the line is equal to the
distance traveled by the pulsar in 1000~years.  A ``kink'' in the bow
shock is labeled with a ``K''; the filament immediately to the north of
the ``K'' is most likely not related to the pulsar.}
\label{fig_pwn}
\end{figure}

Figure~\ref{fig_pwn} shows a smoothed \Ha\ image of the
field surrounding PSR~\psr.  The
position of the pulsar at epoch MJD~52115 is marked, as is the pulsar's
direction of motion with respect to its local standard of rest. The
pulsar is located just inside the apex of an optical nebula with a
clear bow-shock morphology, while the pulsar's projected velocity
points within $\sim30^\circ$ of the apparent nebular  symmetry axis.
This confirms that the nebula is a bow-shock PWN powered by the
pulsar.

The morphology of this PWN deviates in several ways from the classic
bow-shock shape predicted by theory. The eastern half of the nebula is
dominated by a marked kink (indicated with a ``K'' in
Figure~\ref{fig_pwn}), located about $25''$ north-east of the pulsar.
North of this kink the opening angle of the bow shock becomes
significantly broader; this eastern limb of the nebula fades into the
background $\approx1\farcm5$ from the pulsar.  The western half of the
nebula is generally more uniform in its shape and brightness than the
eastern half. On the western side, the opening angle of the bow shock
reaches a maximum $\sim1'$ from the pulsar, before starting to closing
back in on itself. This half of the nebula extends $\sim2'$
from the pulsar.

\Ha\ emission from the head of the nebula is clearly resolved, with a
thickness adjacent to the pulsar of $3''$, decreasing to $2''$ in parts
of the nebula to the pulsar's north-east and north-west.  The head of
the PWN is highly asymmetric around the pulsar's velocity vector, being
considerably broader to the west of this axis than to the east.

The \Ha\ flux density was calculated from an unsmoothed frame in the
manner described by Jones \etal\ (2002\nocite{jsg02}), using the
planetary nebula G321.3--16.7 as a flux standard.  Considering
an $8''$-wide curved strip enclosing the entire visible nebula,
we find that the
\Ha\ flux for the PWN powered by PSR~\psr\ is $2.5 \times
10^{-4}$~\photunits. The mean surface brightness is $7.4 \times
10^{-19}$~\lineSB, three times that measured for PSR~B0740--28
(Jones \etal\ 2002\nocite{jsg02}) or one-tenth of that for PSR~J0437--4715
(\cite{bbm+95}).  Both in the immediate vicinity of the pulsar and at
the kink, the nebula's surface brightness is  $1.5 \times
10^{-18}$~\lineSB, a factor of $\sim2$ brighter than the nebular
average.

\section{Discussion}

The position, speed and direction of motion of PSR~\psr\ have all been
determined to reasonably high precision (see \S\ref{sec_intro}).  This
system thus presents an excellent opportunity to model a bow-shock PWN
and to constrain its parameters.

To quantify the properties of this nebula, we consider
a pulsar of space velocity $V$ moving through an ambient
medium of mass density $\rho_0$ and number density $n_0$.
We define the ``stand-off
distance'', $R_0$, to be the characteristic scale size of the nebula.
For an idealized bow shock, $R_0$ is the distance from the pulsar to the
apex of the bow shock along the direction of motion. The corresponding
``stand-off angle'' is $\theta_0 = R_0 \sin i/D$, where $i$ is the
inclination of the pulsar's motion to the line-of-sight
(see Equation~[4] of \cite{cc02}). We further
define the ``bow-shock angle'', $\psi_0$, to be the angular separation
between the pulsar and the outermost edge of the projected \Ha\ emission
along the projected direction of the pulsar's motion. In general, $\psi_0$
is the only parameter which can be directly measured from an image;
in this case $\psi_0 \approx 2\farcs6$.

We crudely estimate the parameters of the system by assuming that
$i = 90^\circ$, and that the outer edge of the observed \Ha\ emission
demarcates the bow shock structure. In this case $\theta_0 \equiv \psi_0$
and so $R_0 \approx 0.003d_0$~pc.
At the stand-off distance, pressure from the pulsar
wind, $P_{\rm PSR} = \dot{E}/4\pi R_0^2 c$, balances the ram pressure of
the ambient medium, $P_{\rm ISM} = \rho_0 V^2$,
so that  $n_0 = 1.6d_0^{-4}$~cm$^{-3}$.  This confirms
that the pulsar is embedded in the warm, neutral component of
the ISM (e.g.\ \cite{kh88b}), 
as expected for observable \Ha\ bow shocks. The corresponding
ISM sound speed is $c_s \sim5$~\kms, and the Mach number for the pulsar is
$M = V/c_s \ga 12$. 

We now consider the detailed morphology of the head of the nebula. We
first rotate the image counterclockwise by $61^\circ$, so that the
pulsar's motion is strictly from left to right, and adopt an $(x,y)$
coordinate system with the pulsar at the origin.  We then by hand
identify a series of points which demarcate the outermost boundary of
the head of the optical nebula, over an azimuthal range of
$\sim120^\circ$ on either side of the pulsar's direction of motion. The
resulting data-points can then be fit with a variety of functional
forms.  We note that variations in thickness and brightness seen in the
nebula potentially provide additional information on the pulsar's
interaction with its environment. However, to use these data, one must
take into account the width and surface density of the emitting regions as
a function of position, and also must consider the effects of
charge-exchange and collisional excitation (\cite{bb01};
\cite{buc02b}). Given these considerable complications, here we
consider only the overall shape of the emitting regions, as delineated
by the outermost edge of the \Ha\ nebula.

We first consider the idealized case of an isotropic wind propagating
through a homogeneous ambient medium. In this case, the shape of the
bow-shock surface is given by (\cite{wil96}):
\begin{equation}
R(\phi) = \kappa R_0/ \sin\phi \sqrt{3(1-\phi/\tan \phi)},
\label{eqn_axi}
\end{equation}
where $R(\phi)$ is the distance from the pulsar to a given point on the
bow-shock surface, $\phi$ is the angle between the velocity vector and
the vector joining the pulsar to this point, and $\kappa$ is a constant,
with value $\kappa=1$ in this case.

\begin{figure}
\centerline{\psfig{file=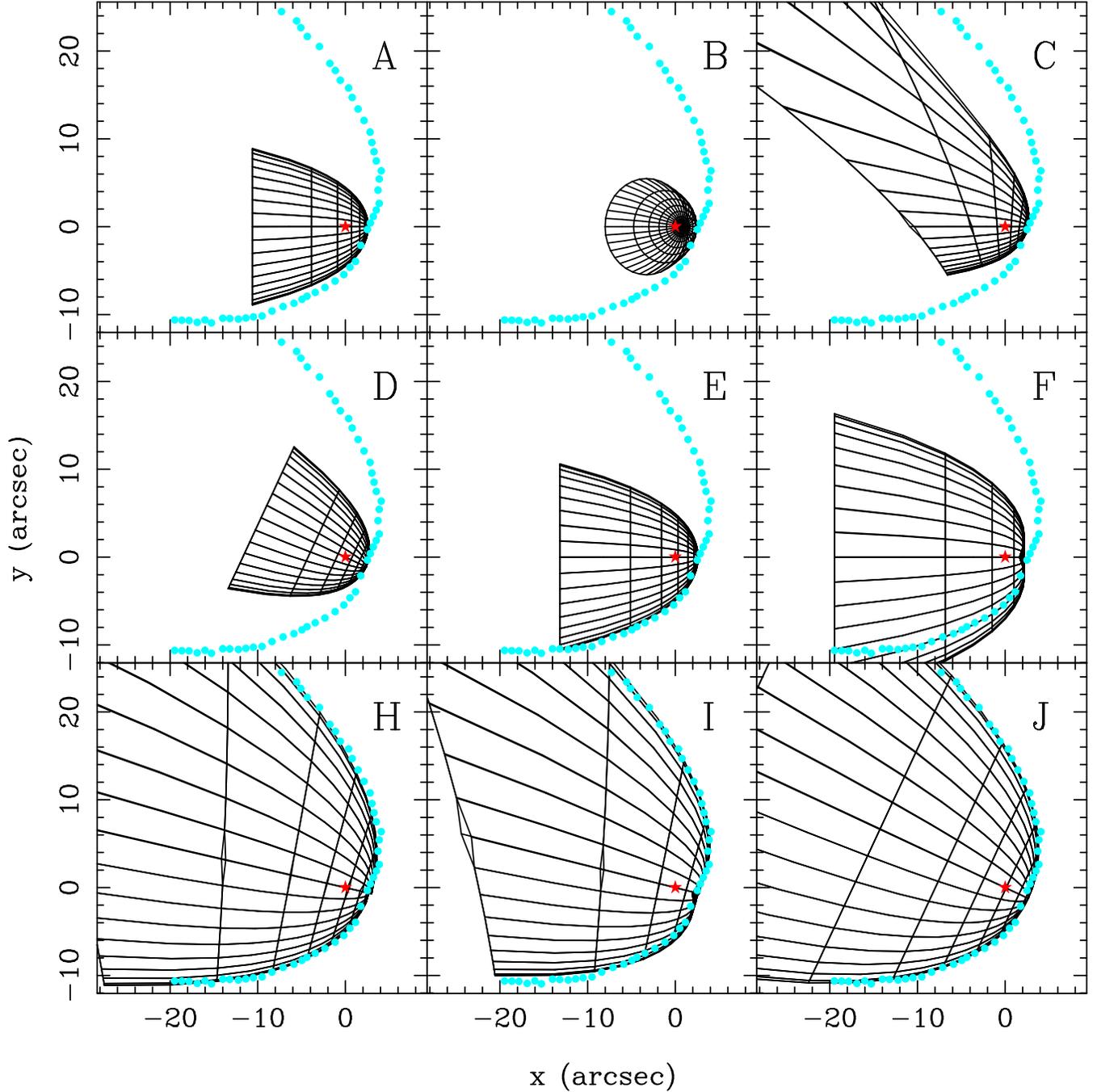,height=18cm,angle=270}}
\caption{Models of the bow-shock nebula powered by PSR~\psr. The nine
panels correspond to nine of the models listed in Table~\ref{tab_models}.
In each panel, the data-points (blue in the on-line version)
represent the outer profile of the
observed bow shock, the star (red in the on-line version)
indicates the position of the pulsar,
and the mesh corresponds to the surface of the model bow shock.
In each case, the model bow shock solutions
are plotted over the range $-140^\circ \le \phi \le 140^\circ$.}
\label{fig_models}
\end{figure}

In panels A and B of Figure~\ref{fig_models}, we show solutions to
Equation~(\ref{eqn_axi}) for $i=90^\circ$ and $i=30^\circ$
respectively, scaled to match the data so that for $\phi = 0^\circ$,
the angular separation between the pulsar and the outermost edge of the
bow-shock surface is $\psi_0 \approx 2\farcs6$.  Best-fit parameters
for each solution are listed in Table~\ref{tab_models}.  (As an aside,
we note that unless $i = 90^\circ$, the usual assumption that $\theta_0
\equiv \psi_0$ is incorrect.  For example, for the inclined case
$i=30^\circ$ we find that $\theta_0 = \kappa R_0\sin i/D = 0\farcs8 \ll
\psi_0$. This indicates that in cases where the inclination angle is
not known, the first-order assumption that $\theta_0 = \psi_0 = \kappa
R_0/D$ results in an {\em upper limit}\ on $\kappa R_0$ and a {\em
lower limit}\ on $n_0$, in contrast to previous calculations [see
e.g.\ \S5.1 of \cite{cc02}].)

These plots make clear that for any inclination, the observed nebula is
too broad and too asymmetric
to be described by Equation~(\ref{eqn_axi}).  However, in
comparing a pulsar bow shock to the shape described by
Equation~(\ref{eqn_axi}), a number of caveats must be made. First,
Equation~(\ref{eqn_axi}) describes a system in which radiative cooling
is efficient, and in which there is significant mixing between the
stellar wind and ambient material (\cite{wil96}), neither of which is
likely to be the case for pulsar bow shocks (\cite{bb01}).  However,
these effects do not change the shape of the bow shock, but simply
require $\kappa \sim 1.7-2.4$ in Equation~(\ref{eqn_axi})
(\cite{buc02a}; \cite{vag+02}).  van der Swaluw \etal\ (2002\nocite{vag+02})
also demonstrate that for low Mach numbers $M \la 5$, the tail of the
resulting bow shock is slightly broader than that described by
Equation~(\ref{eqn_axi}). However, in this case we have demonstrated
above that $M > 12$, sufficiently high that a morphology close to the
analytic shape should result.

We are thus forced to conclude that the idealized bow-shock morphology is
unable to describe this nebula.  To account for the observed appearance,
we must relax some of the assumptions from which Equation~(\ref{eqn_axi})
was derived. Specifically, we need to consider the possibilities of
a density gradient in the ISM, a significant flow velocity for the
ambient medium, or of anisotropies in the pulsar wind.
We briefly summarize how each of these effects manifest
themselves in the observed nebular morphology:

\begin{enumerate}

\item A local density gradient perpendicular to the pulsar's direction
of motion will distort the bow-shock shape of Equation~(\ref{eqn_axi})
so that the head of the nebula is narrower on the side where the
density is higher, and broader on the side where the density is lower.
We here consider a nebula in an exponential density gradient of
e-folding scale $\kappa l R_0$, using the analytic solution provided by
Wilkin (2000\nocite{wil00}).

\item  A density gradient parallel to the pulsar velocity vector
results in a time-variable ambient gas pressure. The stand-off distance
will thus continually change to adjust for this pressure imbalance,
decoupling the position of the pulsar from that of its bow-shock. A
detailed treatment of this process is beyond the scope of this paper;
here we approximate  the effects of such a gradient by using
Equation~(\ref{eqn_axi}), but with the nebula shifted along the $x$-axis
by an amount $-R_1$.  This crudely corresponds to an increase in
ambient density by a factor $\sim(\kappa R_0)^2/(\kappa R_0-R_1)^2$
over a length scale $\sim\kappa R_0$. If we assume an exponential
density gradient with e-folding scale $\kappa m R_0$, we then find that
$m \sim -1/[2\ln (1-R_1/\kappa R_0)]$.  Note that this approximate
treatment only describes the behavior in the narrow region between the
pulsar and the apex of the bow shock; we assume here that the pulsar
has only recently encountered the density gradient, so that the overall
nebular morphology has the form of Equation~(\ref{eqn_axi}) with
constant $R_0$.

\item  Bulk motion of the ambient gas (in the reference frame of
Galactic rotation at that position) will result in a nebula as
described by Equation~(\ref{eqn_axi}), but with the nebula rotated by
some angle $\omega$ about the pulsar position. The angle $\omega$ is
determined by subtraction of the ISM flow velocity vector from the
pulsar velocity vector.

\item An anisotropic wind can generate a wide variety of complicated
bow-shock morphologies, depending on the geometry and orientation of
the outflow (\cite{ban93}; \cite{wil00}).  Wilkin (2000\nocite{wil00})
derives analytic solutions for the bow shocks generated by axisymmetric
winds, parameterizing the resulting morphologies via the variable $c_2$
and $\lambda$. The parameter $c_2$ indicates the nature of the wind
(see Equation~[109] of \cite{wil00}): $c_2 = 3$ corresponds to a
completely polar wind, $c_2 = 0$ to an isotropic wind, and $c_2 = -1.5$
to a completely equatorial wind. The parameter $\lambda$ is the angle
between the pulsar spin axis and the pulsar velocity vector.

\end{enumerate}

We have generated an exhaustive set of bow-shock morphologies
corresponding to each of the above four models, for all reasonable free
parameters and for all inclination angles. For each of these models
alone, we are unable to find a set of parameters which can match the
observed bow-shock morphology:  
a perpendicular density gradient or ISM bulk flow can generate a nebula
which is asymmetric about the $x$-axis (as seen in panels C and D of
Figure~\ref{fig_models}) but which is too narrow to match the
data, while a a parallel gradient or anisotropic wind
can generate a broader bow shock (as seen in panels E and F of
Figure~\ref{fig_models}) but cannot account for the high level of
asymmetry between the top and bottom halves of the nebula.

However, when we combine two or more of these effects, we find that we
can match the morphology
of the head of the nebula with a wide variety of parameters. 
A representative selection of such fits are listed in the lower
half of
Table~\ref{tab_models}; in all cases, the predicted bow-shock
morphology corresponds well to that 
of the \Ha-emitting nebula.
We plot three of these models in Figure~\ref{fig_models}: in model H,
there is a density gradient at an angle to the
pulsar's motion (in this case at an approximate PA of $225^\circ$
in Figure~\ref{fig_models}) and a bulk flow in the ISM; in
model~I, a pulsar with an equatorial wind (with $\lambda = 0^\circ$)
moves through a perpendicular density gradient and a bulk flow; in model J, 
there is both a parallel density gradient and a bulk flow.  This
latter case is particularly noteworthy, as the resultant model is
completely equivalent to that described by Equation~(\ref{eqn_axi}),
but with an apparent pulsar position and direction of motion
which differ from their true values. This makes
clear that fitting to the morphology of the bow shock without knowing
these parameters can give misleading results.

The models which we have invoked to describe the nebular morphology all
correspond to physically reasonable scenarios. Most notably, each of
the successful fits to the data requires both a peculiar flow velocity
for the ambient ISM of magnitude $\sim15-25$~\kms, and an increase in
ambient density of magnitude $\Delta n_0 \sim 1-10$~cm$^{-3}$ over a
scale of $\sim0.02$~pc.  The implied flow velocities are all within the
range of random motions seen in \HI\  clouds (\cite{ars84};
\cite{kf85}), while the required density variations are common in the
warm ISM, being comparable to the typical level of turbulent
fluctuations at these scales inferred from \HI\ absorption experiments
(\cite{des00}). Such density gradients have also been proposed to
account for the morphologies of the bow-shock PWNe powered by
PSRs~B0740--28 and B2224+65 (Jones \etal\ 2002\nocite{jsg02};
\cite{cc02}), and may also account for the ``kink'' seen in
Figure~\ref{fig_pwn}.  Models I, K and L require a pulsar wind which is
dominated  either by an equatorial flow or by polar jets, properties
which have both been seen in many recent PWN observations
(e.g.\ \cite{hss+95}; \cite{gak+02}).  Finally, the mean density
implied by all these models is $n_0 =0.8-1.3$~cm$^{-3}$ (assuming
$\kappa=2$ and $d_0 = 1$), quite typical of the warm, neutral component
of the ISM in which observable \Ha\ pulsar bow-shocks will be
generated.

Given the density gradients inferred for the ambient medium, it is
perhaps surprising that large variations in the H$\alpha$ surface
brightness are not seen as a function of position around the PWN. A
considerably more detailed investigation, incorporating the effects
discussed by Bucciantini \& Bandiera (2001\nocite{bb01}), is needed
to establish whether the ambient density can be quantitatively
probed using the variations in nebular brightness.

\section{Conclusions}

We have identified a striking \Ha\ bow shock powered by the millisecond
pulsar PSR~\psr. The morphology of the nebula does not match the
standard bow shock shape, but can be described by models which include
a significant density gradient in the ISM and a bulk flow of ambient gas;
an anisotropy in the pulsar wind may also contribute to the nebular
morphology. We conclude that in general, it is not possible to uniquely
extract the physical conditions of the system by fitting to the
morphology of the nebula with respect to the central source, nor can
the position and direction of motion of the pulsar be determined from
the shape of the bow shock alone. Rather, we find that additional
information is needed to break the degeneracy between different
possibilities.  In the case of PSR~\psr, a significant flow
velocity for ambient gas can be identified through a non-zero systemic
velocity in optical spectroscopy, a density
gradient along the pulsar's direction of motion can be demonstrated
through relative motion of the pulsar and its bow shock in multi-epoch
imaging, and anisotropies in the pulsar's relativistic
wind may be revealed through high-resolution imaging of the X-ray
emission from \psr\ (e.g.\ \cite{skt+01}) using the {\em Chandra
X-ray Observatory}.  When combined with detailed modeling of the
brightness and width of the nebula as a function of position, these
data can be used to fully characterize the interaction of this pulsar
with its environment.

\begin{acknowledgements}

We thank Matthew Bailes for providing updated timing parameters for
PSR~\psr\ and Shami Chatterjee for helpful discussions. We are
particularly grateful to the referee, Rino Bandiera, for pointing out a
crucial error in our original calculations.  We acknowledge the NTT
team for their assistance.  These results are based on observations
made with ESO Telescopes at the La Silla Observatory (program
67.D--0064).

\end{acknowledgements}


\vspace{2cm}

\scriptsize

\begin{table}[htb]
\caption{Model fits to the bow shock powered by PSR~\psr.  Entries above
and below the horizontal line correspond to models which are unable
to fit and which can successfully fit the observed nebular morphology,
respectively.}
\label{tab_models}
\begin{tabular}{lcc|cccccccl} \hline
Model & $\kappa R_0/D$ & $n_0d_0^4/\kappa^2$ &
  $i$ & $l$ & $R_1/D$ & $\omega$ & $c_2$ & $\lambda$ & Fit? & Comments \\
     & (arcsec) &
 (cm$^{-3}$) & (deg) & & (arcsec) & (deg) & & (deg) \\ \hline
A & 2.6 &  1.6 & 90 & $\ldots$ & $\ldots$ & $\ldots$ & $\ldots$ & $\ldots$ 
& N & Reference case \\
B & 1.6 &  1.0 & 30 & $\ldots$ & $\ldots$ & $\ldots$ & $\ldots$ & 
$\ldots$ & N & Highly inclined \\
C & 2.6 & 1.6 & 90 & 1.2 & $\ldots$ & $\ldots$ & $\ldots$ & $\ldots$
 & N & Perpendicular density gradient \\
D & 2.6 &  1.6 & 90 & $\ldots$ & $\ldots$ & --25 & $\ldots$ & $\ldots$
 & N & ISM bulk flow \\
E & 3.1 & 1.1 &90 & $\ldots$ & 0.5 & $\ldots$ & $\ldots$ & $\ldots$
 & N & Parallel density gradient near apex \\
F & 4.6 & 0.50  & 90 & $\ldots$ & $\ldots$ & $\ldots$ & --1.3 & 0
 & N & Anisotropic wind (mostly equatorial) \\
G & 2.0 & 2.7 & 90 & $\ldots$ & $\ldots$ & $\ldots$ & 1.3 & 0
 & N & Anisotropic wind (mostly polar) \\ \hline
H & 6.9 &  0.22 & 90 & 2.0 & 4.0 & --13 & $\ldots$ & $\ldots$
 & Y & Models C, D and E combined \\
I & 5.8 &  0.32 & 90 & 2.0 & $\ldots$ & --14 &  --1.3 & 0 &
 Y & Models C, D and F combined \\
J & 7.5 &  0.19 & 90 & $\ldots$ & 4.6 & --25 & $\ldots$ & $\ldots$ &
 Y & Models D and E combined  \\ 
K & 7.2 &  0.21 & 90 & 2.0 & 6.5 & --13  & 1.3 & 0 & Y &
Models C, D, E, G combined \\
L & 5.9 &  0.31 & 90 & 2.4 & 4.0 & --12 & --1.3 & 90 & Y &
Models C, D, E, F with $\lambda = 90^\circ$ \\ \hline
\end{tabular}
\end{table}

\normalsize

\clearpage

\end{document}